\documentclass[a4paper,prl,twocolumn]{revtex4-1}
\usepackage{graphicx}
\usepackage{amsmath}
\usepackage{fancyhdr}
\usepackage{cancel}
\usepackage{color}
\setcitestyle{round}
\begin{document}
\title{Detecting temperature fluctuations at equilibrium}
\author{Purushottam D. Dixit}
\affiliation{Department of System Biology, Columbia University}
\thanks{email:pd2447@columbia.edu}
\begin{abstract}
Gibbs and Boltzmann definitions of temperature agree only in the macroscopic limit. The ambiguity in identifying the equilibrium temperature of a finite sized `small' system exchanging energy with a bath is usually understood as a limitation of conventional statistical mechanics. We  interpret this ambiguity as resulting from a stochastically fluctuating temperature coupled with the phase space variables giving rise to a broad temperature distribution.  With this ansatz, we develop the equilibrium statistics and dynamics of small systems. Numerical evidence using an analytically tractable model  shows that the effects of temperature fluctuations can be detected in equilibrium and dynamical properties of the phase space of the small system. Our theory generalizes statistical mechanics to small systems relevant to biophysics and nanotechnology.
\end{abstract}
\maketitle

{\bf Introduction}:  Equilibrium properties of a macroscopic system exchanging energy with a bath can be described by a single intensive paramter, its temperature, with remarkable accuracy; independently of the chemical nature of the bath and system-bath interactions owing to weak coupling between the system and the bath.  On the other hand,  it is unlikely that a bath couples weakly to a system with small number of degrees of freedom; consequently, small systems including biophysical polymers~\citep{pohorille2010good} and nanomagnets~\citep{chamberlin2009beyond} show considerable deviations from the traditional statistical mechanical description~\citep{hill}. Mathematically, no inverse temperature $\beta$ exists such that the exponential canonical ensemble distribution accurately predicts equilibrium properties of a small system solely dependent on its Hamiltonian.  Alternatively, Gibbs' definition of temperature which depends on the {\it typical} value of energy and the Boltzmann's definition of temperature which depends on the {\it mean} value of energy differ substantially from each other in the case of small systems systems~\citep{mandelbrot2008temperature,dunkel2013consistent}.  Traditionally, this ambiguity is interpreted as an inevitable statistical uncertainty in parameter estimation or a limitation of statistical mechanics~\citep{mandelbrot2008temperature,mandelbrot1962role,uffink1999thermodynamic,schlogl1988thermodynamic,van2013temperature}. 

In this communication, instead of treating the ambiguity in identifying a unique temperature as a limitation, we let go of the notion of a unique temperature, especially for small systems. We identify the ambiguity as a consequence of a broad distribution $p_{\rm eq}(\beta)$. Furthermore, we identify the broad temperature distribution  as the $\bar r-$marginalization of the joint equilibrium distribution $p_{\rm eq}(\bar r, \beta)$ of the stochastic variable $(\bar r(t), \beta(t))$ where $\bar r$ is the phase space of the system.  

Using maximum entropy arguments, we first estimate the joint equilibrium distribution $p_{\rm eq}(\bar r, \beta)$ by introducing two new intensive parameters in the hyperensemble. We then show how our theory reduces to traditional statistical mechanics of macroscopic systems in the suitable limit. We  illustrate a connections with non-extensive statistical mechanics of Tsallis~\citep{tsallis1988,wilk2000interpretation} and our theory at thermodynamic equilibrium. Then, we propose Fokker-Planck and Langevin equations for the time evolution of the instantaneous distribution $p(\bar r, \beta; t)$. Finally, using realistic all atom molecular dynamics simulations, we present numerical evidence to support our framework and discuss its limitations.

{\bf Statitical mechanics of small systems}: Consider a small system  as above. Due to possible non-weak coupling between system and the bath, the equilibrium phase space distribution of the system $p_{\rm eq}(\bar r)$ will depend on the nature of system-bath interactions~\citep{dixit:bj11,dixit2011elastic,dixit2012role}. Let us work with the ansatz that  the non-canonical behavior arises because the temperature of the system fluctuates~\citep{wilk2000interpretation,touchette2004temperature,van2013temperature}. The joint equilibrium distribution is simply
\begin{eqnarray}
p_{\rm eq}(\bar r, \beta) = p_{\rm eq}(\bar r|\beta) \times p_{\rm eq}(\beta). \label{eq:pjoint}
\end{eqnarray}

In Eq.~\ref{eq:pjoint}, 
\begin{eqnarray}
p_{\rm eq}(\bar r|\beta) = e^{\beta \left (F(\beta) - H(\bar r) \right )} \label{eq:canon}
\end{eqnarray} is the usual Boltzmann distribution and $p_{\rm eq}(\beta)$ needs to be determined.  Since there are no conservation laws for temperature, Gibbs' ensemble picture is inapplicable. We resort to an equally valid alternative. We employ the maximum entropy (maxEnt) framework~\citep{Press2012,it_theory1}. We maximize the entropy of the joint distribution $p(\bar r,\beta) = p(\bar r|\beta) \times p(\beta)$ subject to suitable constraints.  The entropy of the joint distribution is given by
\begin{eqnarray}
S\left [p(\bar r,\beta) \right ] &=& -\sum_{\bar r, \beta} p(\bar r, \beta) \log p(\bar r, \beta) \\
&=& -\sum_{\beta} p(\beta) \log p(\beta) + \sum_\beta s(\beta) p(\beta) \label{eq:entropy}
\end{eqnarray}
where 
\begin{eqnarray}s(\beta) &=& -\sum_{\bar r} p(\bar r|\beta) \log p(\bar r|\beta), \\
p(\beta)&=& \sum_{\bar r} p(\bar r,\beta)
\end{eqnarray} is the $\bar r-$marginal of $p(\bar r,\beta)$, and $p(\bar r|\beta)$ is given by Eq.~\ref{eq:canon}.


When determining $p_{\rm eq}(\beta)$ the choice of constraints is important. Since the temperature of the system is not fixed, we choose $\langle \beta \rangle$ as a constraint. Also, while the entropy of the composite macroscopic system comprising the system and the surrounding bath is maximized, the entropy of the small system itself not. Consequently, we choose the average entropy $\langle s(\beta) \rangle$ as an additional constraints and maximize $S\left [p(\bar r,\beta) \right ] $ using Lagrange multipliers. The constraint of average entropy is a common in statistical physics and Bayesian statistics of hyperensembles.  See~\citep{caticha2004maximum,dixit2013quantifying,Crooks2008,dixitmaxent} for different motivations behind this choice.  After maximization, we find that the equilibrium distribution $p_{\rm eq}(\beta)$ is estimated by 
\begin{eqnarray}
p_{\rm eq}(\beta) &=& \frac{e^{\lambda s(\beta) -  \zeta \beta }}{\mathcal Z(\lambda, \zeta)} \label{eq:pbeta_general}
\end{eqnarray}
In Eq.~\ref{eq:pbeta_general}, $\mathcal Z$ is a generalized partition function and $\lambda$ and $\zeta$ are Lagrange multipliers that determine the shape of $p_{\rm eq}(\beta)$.  If entropy $s(\beta)$ is a unitless number, then $\lambda$ is unitless and $\zeta$ has the units of $1/\beta$. The physical interpretation of these Lagrange multipliers will become clearer below.

The joint equilibrium distribution $p_{\rm eq}(\bar r,\beta) = p_{\rm eq}(\bar r|\beta) \times p_{\rm eq}(\beta)$ is 
\begin{eqnarray}
p_{\rm eq}(\bar r,\beta) = \frac{e^{\beta F(\beta) - \beta H(\bar r) + \lambda s(\beta) - \zeta \beta}}{\mathcal Z(\lambda, \zeta)}\label{eq:jointrb}
\end{eqnarray}
Thus, instead of describing a thermally equilibrated small system with one intensive parameter, its inverse temperature $\beta$, our framework requires two intensive parameters $\lambda$ and $\zeta$ whose meaning will become clear below.

%

{\bf Connections to traditional statistical mechanics:} Assume that the entropy $s(\beta)$ is monotonically decreasing in $\beta$, a reasonable assumption for systems with monotonically increasing density of states. A straightforward calculation shows that the maximum of $p_{\rm eq}(\beta)$ is situated at $\beta = \beta_0$ where $\beta_0$ is such that $\zeta / \lambda = -c(\beta_0)/\beta_0$. Here, $c(\beta_0)$ is the heat capacity of the system when interacting with an ideal gas at inverse temperature $\beta_0$.  

In the limiting case when  $\lambda \rightarrow \infty$ and $\zeta \rightarrow \infty$ such that their ratio is constant, non-negligible contribution to $p_{\rm eq}(\beta)$ comes only from near $\beta = \beta_0$ and $p_{\rm eq}(\beta) \approx \delta(\beta-\beta_0)$ where $\delta(x)$ is the Dirac Delta function. This is exactly the traditional  canonical ensemble picture where the system is assigned the temperature of the surrounding thermal bath. It is  clear that the magnitudes of $\lambda$ and $\zeta$ dictate the breadth of the $p_{\rm eq}(\beta)$ distribution and hence the deviation from canonical ensemble. The ratio $\lambda/\zeta$ dictates the most likely tempearture of the system.

{\bf Connection to non-extensive statistical mechanics:} Systems that do not obey the conventional distributions from statistical mechanics are sometimes entertained within a framework called non-extensive statistical mechanics~\citep{tsallis1988}. Though not commonly invoked for small systems at equilibrium, here, we will demonstrate that non-extensive statistical mechanics can be arrived at by marginalization over temperature in a hyperensemble.

Consider a system whose entropy scales as logarithm of temperature, $s(\beta) = s_0 \log \beta$, and the internal energy scales proportional to the temperature, $U(\beta) = U_0/\beta$, when coupled to a bath of ideal gas particles at inverse temperature $\beta$. These are excellent assumptions for bound systems where density of states increases monotonically with energy. Examples include ideal gas in a container and a collection of harmonic oscillators. From Eq.~\ref{eq:pbeta_general}, we have
\begin{eqnarray}
p_{\rm eq}(\beta) &=& \frac{e^{-\beta  \zeta } \beta ^{\lambda s_0} \zeta ^{\lambda s_0 +1}}{\Gamma (\lambda s_0 +1)}. \label{eq:pbeta}
\end{eqnarray}
Eq.~\ref{eq:pbeta} is a Gamma distribution also known as the generalized $\chi-$squared distribution. Interestingly, a gamma distributed inverse temperature is very commonly used in a superstatistical explanation of non-extensive statistics~\citep{superstatistics}. Marginalizing over gamma distributed inverse temperature in Eq.~\ref{eq:pjoint} results in the so called ``Tsallis statistics" for the phase space. We have
\begin{eqnarray}
p_{\rm eq}(\bar r,\beta) &=& \frac{e^{\beta \left ( U(\beta) - s(\beta)/\beta \right ) - \beta H(\bar r) + \lambda s(\beta) - \zeta \beta}}{\mathcal Z(\lambda, \zeta)}\\
&=&  \frac{e^{ U_0 - \beta (H(\bar r) + \zeta) + (\lambda-1) s_0 \log(\beta) }}{\mathcal Z(\lambda, \zeta)}.
\end{eqnarray}
Integrateing over $\beta$, we have
\begin{eqnarray}
p_{\rm eq}(\bar r) &\propto& (1 - \beta_0(q-1) H(\bar r) )^{\frac{1}{q-1}}.\label{eq:tsallis}
\end{eqnarray}
Eq.~\ref{eq:tsallis} is the $q-$generalized canonical ensemble distribution in Tsallis statistics where 
\begin{eqnarray}
q=\frac{s_0-\lambda }{s_0 - \lambda-1}~{\rm and}~\beta_0 = \frac{\lambda -s_0+1}{\zeta }.
\end{eqnarray} 

In the framework of non-extensive statistical mechanics, one arrives at Eq.~\ref{eq:tsallis} by maximizing Tsallis' $q$ entropy with respect to $p(\bar r)$ by constraining an unnatural {\it escort expectation of energy}~\citep{tsallis1988}.

In this work, in contrast to deriving $p_{\rm eq}(\bar r)$ by maximizing the non-extensive Tsallis entropy by constraining an unnatural expectation value,  we derive it from a superstatistical distribution Eq.~\ref{eq:jointrb} and additional assumptions about $p_{\rm eq}(\beta)$ and system behavior. In our derivation, the gamma distribution $p_{\rm eq}(\beta)$ arises in a context specific manner i.e. through the logarithmic dependence of the entropy on the inverse temperature and by constraining average inverse temperature.  Therefore, starting from the extensive Gibbs-Shannon entropy, maxEnt can act as a predictive framework for constructing non-extensive {\it effective entropies}~\citep{hanel2011generalized} of which  the Tsallis entropy is a particular example. 

Previously,  non-extensive entropies have been criticized from an Occam's razor point of view~\citep{Press2012,presse2013nonadditive,peterson2013maximum,presse2014nonadditive} when compared to the Gibbs-Shannon entropy. Our work suggests that non-extensive entropies may arise as `effective entropies' when considering extensive entropies in a hyperensemble. Nevertheless, there is a potential loss of information when marginalizing over the temperature $\beta$ in the hyperensemble that is inherent to constructing these effective entropies. We believe that the above demonstration argues in favor the extensive Gibbs-Shannon entropy, albeit in a hyperensemble, even when the observable phase space may show non-extensive behavior.

{\bf  Stochastic Dynamics:} For simplicity of notation, let us consider a one dimensional system. The simplest time evolution of the instantaneous distribution $p(r, \beta;t)$ of the extended phase space that relaxes to  a prescribed equlibrium distribution $p_{\rm eq}(r, \beta)$ can be modeled by an {\it over damped} Smoluchowski equation. We have
\begin{eqnarray}
\frac{\partial p( r, \beta;t)}{\partial t} &=& - \left ( \frac{1}{\gamma_r} \frac{\partial }{\partial r} \left [f_r \cdot  p \right ] + \frac{1}{\gamma_\beta}\frac{\partial }{\partial \beta}  \left [ f_\beta \cdot p \right ]  \right ) \nonumber \\ &+& D_r \frac{\partial^2 p}{\partial r^2} + D_\beta \frac{\partial^2 p}{\partial \beta^2}  \label{eq:fp}
\end{eqnarray}
where the `forces' $f_r$ and $f_\beta$ are defined as
\begin{eqnarray}
f_r = \frac{\partial}{\partial r} \log p_{\rm eq}(r,\beta)~{\rm and}~f_\beta = \frac{\partial}{\partial \beta} \log p_{\rm eq}(r,\beta). 
\end{eqnarray}
By construction, Eq.~\ref{eq:fp} will relax to the equilibrium distribution $p_{\rm eq}(r, \beta)$ {\it if} $D_r  = 1/\gamma_r$ and $D_\beta = 1/ \gamma_\beta$. Note that the statistical properties of $(r(t),\beta(t))$ can also be estimated by an overdamped Langevin equation (Brownian dynamics) that is equivalent to Eq.~\ref{eq:fp}. The  Langevin equation reads 
\begin{eqnarray}
\dot r &=& D_rf_r +\sqrt{2D_r} \eta_r \nonumber \\
\dot \beta &=&D_\beta f_\beta +\sqrt{2D_\beta} \eta_\beta \label{eq:langevin2d}
\end{eqnarray}
Here, $\eta_r$ and $\eta_\beta$ are usual uncorrelated Gaussian random variables with unit variance. 

%
{\bf Linear analysis:} It is instructive to study a linear system before analyzing realistic molecules. Consider a one dimensional harmonic oscillator interacting with a thermal bath. If the deviations from a canonical distribution are negligible, we can treat Eq.~\ref{eq:langevin2d}  in the linear regime by expanding $f_r$ and $f_\beta$ to the first order in $r$ and $\beta$. In the linear approximation, the joint equilibrium distribution $p_{\rm eq}(r, \beta)$ will be described by a joint normal distribution. The simplest coupled system of overdamped Langevin equations for $r(t)$ and $\beta(t)$ that relaxes to to a joint normal distribution is given by
\begin{eqnarray}
\dot r &\approx& l_{11} r + l_{12} \beta + \eta_r \\
\dot \beta &\approx& l_{21} r + l_{22} \beta + \eta_\beta \label{eq:linear}
\end{eqnarray}
We have assumed that the variables $r$ and $\beta$ are appropriately scaled by absorbing the diffusion constants $D_r$ and $D_\beta$,  $l_{ij}$ are the scaled linear expansion coefficients of $f_r$ and $f_\beta$, and $\eta_r$ and $\eta_\beta$ are the usual uncorrelated Gaussian noises. Integrating over $\beta(t)$ and substituting in $\dot r$, we get
\begin{eqnarray}
\dot r &=& l_{11} r + l_{12}e^{l_{22} t}\int_0^t ds \cdot  l_{12}\cdot e^{-l_{22}s} \nonumber\\ &+& l_{12} e^{l_{22}t}\int_0^t ds \cdot \eta_\beta \cdot e^{-l_{22} s} + \eta_r \\
\Rightarrow \ddot  r &=& \left ( l_{11} + l_{22} \right ) \dot r +\left ( l_{12} l_{21} - l_{11} l_{22} \right ) r  \nonumber \\ &+&  \left ( l_{12} \eta_\beta - l_{22} \eta_r \right ) + \dot \eta_r \label{eq:purple}
\end{eqnarray}
The time derivative of white noise $\dot \eta_r$ is a purple noise which has quadratically increasing power spectrum. The dynamics of temperature fluctuations are governed by the linear terms $l_{12}$, $l_{21}$, $l_{22}$, and the white noise $\eta_\beta$. These terms also appear in the effective Langevin equation for $r(t)$. The linear analsysis suggests that one can infer the of dynamics of $\beta(t)$ by observing the dynamics of $r(t)$. 

The dynamics of $r(t)$ is governed by a much richer equation than the usual overdamped Langevin equation.  A one dimensional small linear harmonic oscillator exchanging energy with a thermal bath can be modeled by a second order Langevin equation with a combination of white and purple noise. These predictions can be tested by observing dynamical properties of a small colloidal particle trapped in a harmonic well using optical traps. 

%


{\bf A `small' harmonic oscillator:} 
How do we verify the effects of temperature fluctuations on the phase space of a small system? We resort to realistic molecular dynamics simulations of an analytically tractable system viz. a harmonic oscillator.

Consider a three dimensional dumbell shaped Lennard-Jones harmonic oscillator interacting non-weakly with a bath. Realistic examples include colloidal beads tied to each other by a biopolymer or linear molecules such as CO$_2$.  The canonical ensemble distribution for the Harmonic oscillator is given by
\begin{eqnarray}
p_{\rm eq}(r|\beta) = \frac{4 \beta ^{3/2} r^2}{\sqrt{\pi }} \times e^{-\beta  r^2} \label{eq:hcanon}
\end{eqnarray}
where $r$ is the displacement of the oscillator. Without loss of generality, we have assumed that the spring constant of the oscillator is $k=2$. If the system-bath interactions are non-negligible, we expect that the equilibrium phase space distribution of the oscillator will deviate considerably from the Boltzmann distribution.

The entropy of the oscillator scales as $s(\beta) \sim \log \beta$ and from Eq.~\ref{eq:pbeta_general}, we know that the equilibrium distribution $p_{\rm eq}(\beta)$ will be governed by a Gamma distribution 
\begin{eqnarray}
p_{\rm eq}(\beta) &=& \frac{e^{-\beta  \zeta } \beta ^{\lambda } \zeta ^{\lambda  +1}}{\Gamma (\lambda +1)}. \label{eq:pbeta}
\end{eqnarray}
The joint equilibrium distribution $p_{\rm eq}(r,\beta) = p_{\rm eq}(r|\beta) \times p_{\rm eq}(\beta)$ on the other hand is obtained by multiplying Eq.~\ref{eq:hcanon} and Eq.~\ref{eq:pbeta}
\begin{eqnarray}
p_{\rm eq}(r, \beta) &=& \frac{4 r^2 \beta ^{\lambda +\frac{3}{2}} \zeta ^{\lambda +1} e^{-\beta  \left(\zeta +r^2\right)}}{\sqrt{\pi } \Gamma (\lambda +1)}. \label{eq:joint}
\end{eqnarray}
Integrating over all values of $\beta$, we obtain the marginal $r$ distribution
\begin{eqnarray}
p_{\rm eq}(r) = \frac{4 r^2 \zeta ^{\lambda +1} \Gamma \left(\lambda +\frac{5}{2}\right) \left(\zeta +r^2\right)^{-\lambda -\frac{5}{2}}}{\sqrt{\pi } \Gamma
   (\lambda +1)}. \label{eq:prmarginal}
\end{eqnarray}

Moreover, we can also model the dynamics of the oscillator by the coupled Langevin equation of Eq.~\ref{eq:langevin2d}. From Eq.~\ref{eq:joint}, the ``forces'' $f_r$ and $f_\beta$ are given by
\begin{eqnarray}
f_r  &=& \frac{2}{r} - 2r \beta~{\rm and}~f_\beta = \frac{3 - 2\beta r^2 - 2\beta \zeta + 2\lambda}{2\beta} \label{eq:hforce}
\end{eqnarray}

Eq.~\ref{eq:prmarginal} along with Eq.~\ref{eq:langevin2d} where the forces $f_r$ and $f_\beta$ are given by Eq.~\ref{eq:hforce} are our predictions for the Harmonic oscillator regardless of the bath that is interacting with. These predictions can be tested experimentally or in a realistic numerical simulation.

\begin{center}
\begin{figure}
\includegraphics[scale=0.65]{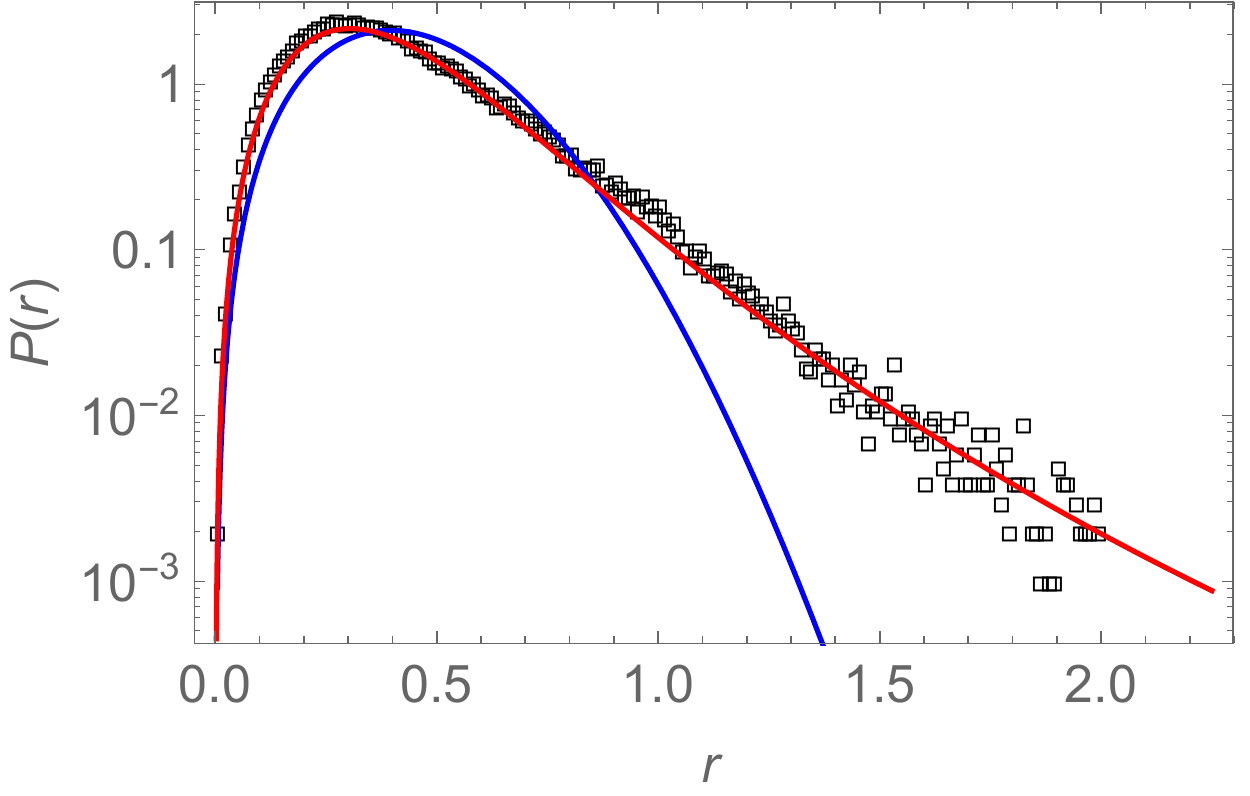}
\caption{We study the equilibrium properties of a 3D dumbbell shaped harmonic oscillator comprising of Lennard Jones particles interacting with a bath of water molecules at 300 K (see appendix I for details) using all atom MD simulations. The numerically obtained marginal distribution $p(r)$ of the oscillator separation $r$ (black squares) is better captured by Eq.~\ref{eq:prmarginal} (red line)  than the usual canonical ensemble distribution of Eq.~\ref{eq:hcanon} (blue line). \label{fg:fitpr}}
\end{figure}
\end{center}

{\bf Numerical validation:} With the aid of MD simulations of a dumbbell shaped Lennard-Jones harmonic oscillator coupled to a bath of water molecules at 300 K (see appendix I for details), we confirmed the numerical superiority of Eq.~\ref{eq:prmarginal} compared to Eq.~\ref{eq:hcanon} and estimated the parameters $\lambda\approx 2.19$ and $\zeta\approx 0.34$. Fig.~\ref{fg:fitpr} shows that Eq.~\ref{eq:prmarginal} which allows for a broad temperature distribution indeed fits the numerically estimated distribution much better than the usual canonical ensemble distribution of Eq.~\ref{eq:hcanon}. It is clear that by allowing the inverse temperature to have a broad distribution,  the equilibrium properties of the harmonic oscillator interacting with its thermal surroundings are captured correctly.

The dynamics of $r(t)$ can be predicted using Eq.~\ref{eq:fp} by studying the equivalent Langevin equation (see appendix II). In Fig.~\ref{fg:autocorr} we compare the numerically estimated autocorrelation function
\begin{eqnarray}
C(\tau) = \langle r(\tau) r(0) \rangle_{\rm eq} - \langle r \rangle_{\rm eq}^2
\end{eqnarray}  from  MD simulation (black squares) and the prediction from the 2-d Langevin equation (red). The predictions from an analogous 1-d Langevin equation that relaxes to $p_{\rm eq}(r)$ of Eq.~\ref{eq:prmarginal} are shown in blue.  While the dynamics observed in the MD simulation has two time scales resulting in a double exponential decay in the autocorrelation function, the 1-d Langevin equation is only able to capture one effective time scale. On the other hand, the 2-d Langevin equation has two natural time scales governed by $D_r$ and $D_\beta$ respectively. The coupled Langevin equation equivalent to Eq.~\ref{eq:fp} (see appendix II) with $dt =5\times 10^{-8}$ and $D_\beta \approx 50 \times D_r$ does indeed captures the autocorrelation function while an analogous 1-d equation fails to do so  (see appendix II for details of the fit). 

In appendix III we show that the theoretical predictions are valid over a range of bath temperatures and system-bath interactions. In this work, we  study a system whose canonical ensemble distribution can be analytically computed and the entropy analytically estimated. This allowed us to compute $p_{\rm eq}(\beta)$ and $p_{\rm eq}(r, \beta)$ analytically. For more realistic systems with multiple degrees of freedom, $p_{\rm eq}(r|\beta)$ needs to be estimated numerically along with $p_{\rm eq}(\beta)$.

In summary, the mesoscopic harmonic oscillator interacting with a thermal bath of water molecules shows significant deviation from the canonical ensemble description. We can correctly predict both equilibrium {\it and} dynamical properties of the oscillator by allowing its temperature to vary as a stochastic variable which is coupled with the phase space variable $r(t)$. 
\begin{center}
\begin{figure}
        \includegraphics[scale=0.66]{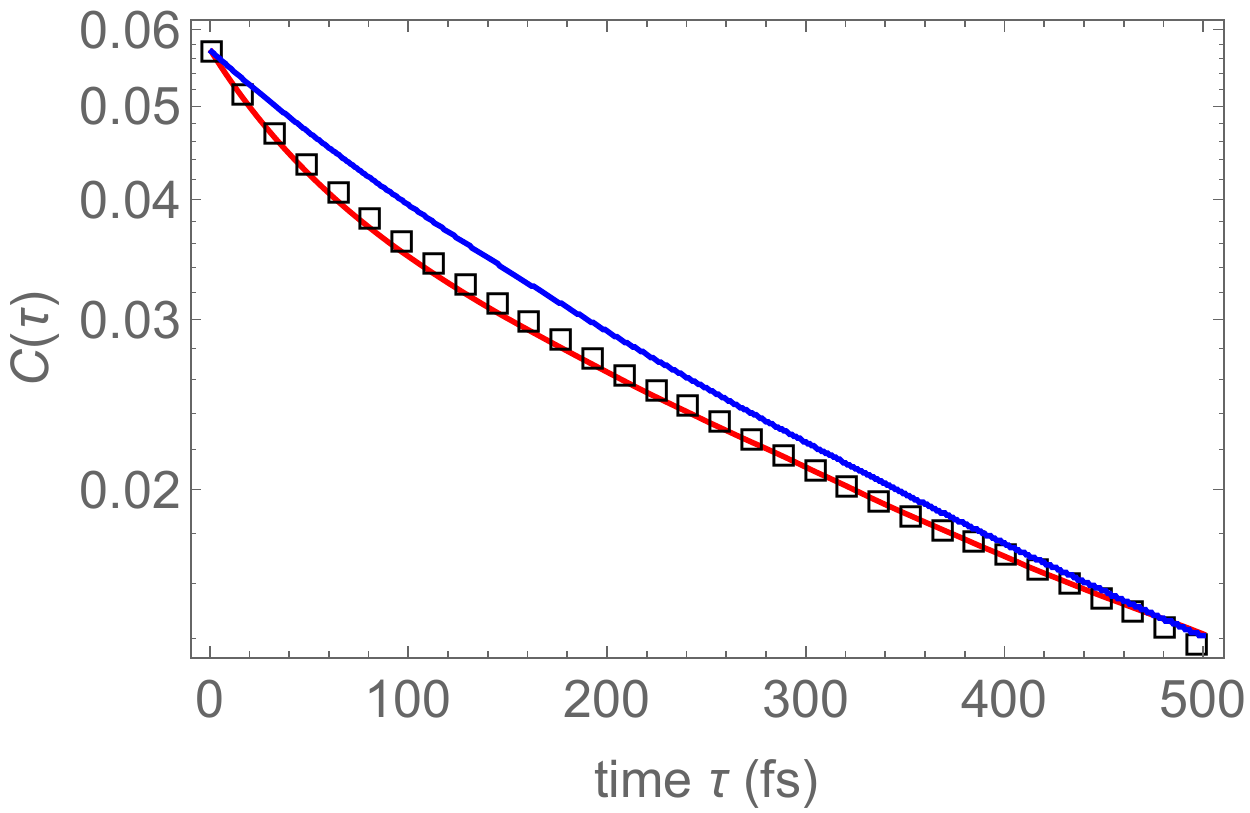}
        \caption{We study the autocorrelation function of the harmonic oscillator interacting with a bath of water molecules. We model the dynamics of the extended phase space $(r(t), \beta(t))$ using a simple coupled Langevin equation (see appendix II). We find that the 2 dimensional Langevin equation (red line) captures the two time scales inherent to the dynamics of $r(t)$ as observed in MD simulations (black squares). On the other hand, an analogous 1-d Langevin equation can only capture one effective time scale (blue line).\label{fg:autocorr}}
\end{figure}
\end{center}

{\bf Discussion:}   

It is known that, at equilibrium, mesoscopic systems have larger fluctuations compared to a macrosopic system. We have argued that these enhanced fluctuations be understood as arising from a dynamically fluctuating temperature. 

How do we reconcile a time dependent temperature, a non-equilibrium phenomena {\it prima facie}, in an equilibrium setting? Even though the temperature is changing, the extended phase space $(\bar r(t), \beta(t))$ is still governed by a detailed-balanced Markov process. It's an easy calculation to show that the entropy production, as defined in stochastic thermodynamics~\citep{seifert2008stochastic}, is indeed zero for the hyperensemble. Nevertheless, there are multiple questions which need resolution. For example,   How do we formulate non-equilibrium phenomena in the hyperensemble setting? For example, how do we modify non-equilibrium fluctuation relationships~\citep{jarzynski2013equalities} for small systems?  We leave this to future work.

\acknowledgments{{\bf Acknowledgment:} We thank Ken Dill, Steve Press\'e, Dilip Asthagiri for a critical reading of the manuscript. We thank Sumedh Risbud, Karthik Shekhar, Manas Raach, and Anjor Kanekar for fruitful discussions.}


\section{Appendix I: MD simulations}
A harmonic dumbbell oscillator consisting of two Lennard-Jones particles was immersed in a bath of 333 TIP3~\citep{tip32,tip3mod} water molecules. NVT molecular dynamics simulations were run with NAMD~\citep{namd:cc05} at 300K with a box size of 19.12\AA. The CHARMM~\citep{charmm:jpcb98} forcefield was used to describe the interaction between the harmonic oscillator particles and surrounding water molecules. The spring constant for the dumbell was chosen to be $k=0.25$~kcal/mol$\cdot$\AA$^2$,  the $\epsilon$ parameter was set at $\epsilon=-20.0$ kcal/mole and the size parameter was set at $r = 1$\AA. The systems were minimized for 2000 steps followed by an equilibration of 1 nanosecond and a production run of 2 nanosecond. The integration time step was 0.25 femtoseconds and the trajectory was saved every 2.5 femtoseconds.

\section{Appendix II: Fitting Langevin dynamics to data}
The coupled Langevin equation corresponding to Eq.~\ref{eq:fp} where the equilibrium distribution $p_{\rm eq}(r, \beta)$ is given by Eq.~\ref{eq:pbeta} is given by
\begin{eqnarray}
 \left ( \begin{array}{c} r(t+dt) \\  \beta(t+dt) \end{array}\right ) &\approx&  \left ( \begin{array}{c} r(t) \\  \beta(t) \end{array}\right ) + dt \left ( \begin{array}{c} D_r f_r \\ D_\beta f_\beta \end{array}\right ) \nonumber \\&+& \sqrt{2dt} \left ( \begin{array}{c} \eta_r \sqrt{D_r} \\  \eta_\beta \sqrt{D_\beta} \end{array}\right ) \label{eq:langevin2dn} 
\end{eqnarray}
Here, $\eta_r$ and $\eta_\beta$ are uncorrelated Gaussian random variables with unit variance, $dt$ is a small time step, $D_r$ and $D_\beta$ are diffusion coefficients for the phase space coordinate $r$ and the temperature $\beta$.

From the MD simulation, we first estimated the autocorrelaion function $C(\tau)$. The Langevin equation can be scaled in time by multiplying the diffusion constants and dividing the time step $dt$ by the same number. In order to ensure smooth integration, we first set the integration time step to a very small value; $dt = 5\times 10^{-8}$. Every pair $(D_r, D_\beta)$ of diffusion constants predicted an autocorrelation function that had two inherent time scales manifested in a double exponential decay.  We manually scanned the $(D_r, D_\beta)-$space  to match the MD-autocorrelation function. We found that $D_r = 1$ and $D_\beta = 50$ gave reasonable fits (red curve).

We also wrote down a 1-d Langevin equation analogous to Eq.~\ref{eq:langevin2dn},
\begin{eqnarray}
r(t+dt) \approx r(t) + D_r f_r dt + \sqrt{2D_r dt}\eta_r
\end{eqnarray}
where $f_r = \frac{d}{dr} \log p_{\rm eq} (r)$ (see Eq.~\ref{eq:prmarginal}). This equation had only one diffusion constant $D_r$. A one dimensional scan of $D_r$ suggested that the autocorrelation function predicted using the 1-d Langevin equation always had a single exponential decay. We found the best fit to the autocorrelation function at $D_r \approx 50$ (blue curve).

\end{document}